\documentstyle[12pt]{article}
\title{On the electrodynamics of a magnetized rotating accretor}
\author{Aldo Treves and Ewa Szuszkiewicz \\
                                          \\
{\it International School for Advanced Studies, SISSA} \\
{\it via Beirut 2-4, 34013 Trieste, Italy}}
\date{}

\begin{document}

\vspace{4.cm}
\maketitle

\vspace{2.cm}
{\bf Summary}

We consider the current flow in an accreting rotating neutron star (X-ray
pulsator) and propose that the active magnetosphere is limited by the
Alfv\`en surface.
Formulae for the luminosity and torque due to currents circulating in
the neutron star are given, and astrophysical consequences for systems
like Her X-1 and LMXRB are shortly discussed.

\vspace{2.cm}
\begin{center}
SISSA Ref. 31/95/A (March 1995)
\end{center}
\vspace{2.cm}
\begin{center}
Accepted by Monthly Notices of the Royal Astronomical Society
\end{center}
\newpage

\section{Introduction}

The electrodynamics of an isolated rotating magnetic neutron star
is the backbone of pulsar models. The  fundamentals of the theory have
been posed in the paper of Goldreich and Julian (1969). The complete
problem has been recognized to be  complicated requiring the use
of rather advanced techniques of plasma physics.

The other astrophysical situation where one deals with magnetized
neutron stars is that of X-ray pulsators, where the neutron star
accretes matter from a close companion. The basis of the theory was
formulated in the seventies, and its capacity to account for the spin
up (down) of X-ray pulsators can be considered as a test of its
validity (e.g. Lipunov 1992, and references therein).

An aspect which has not focused much attention is that of the study
of the electrodynamic activity, powered by the rotation of the
neutron star (\`{a} la pulsar), when the neutron star is surrounded by
a plasma sphere due to the very accretion process. This is the
argument which we consider here in  a rather simplified way.

In section 2 we briefly summarize the basics of pulsar electrodynamics.
In section 3 we recall some formulae relevant for describing accretion
on a static magnetic neutron star. Our elementary formulation for
the electrodynamics of a magnetic rotating accretor is proposed
in section 4. Astrophysical applications with consideration to
Her X-1 and  of a representative case for low mass X-ray
binaries are discussed in section 5.

\section{ Pulsar electrodynamics}

The aligned pulsar - unipolar inductor
analogue  - was proposed in the
seminal paper of Goldreich and Julian (1969). In particular it was
recognized that the magnetosphere intrinsically divides into an
equatorial inactive part, which corotates with the pulsar, and a
polar part 
with current flowing along the field lines.
 The colatitude at the surface of the star (with the radius $a$)
 of the magnetic surface separating the two zones is given in the
approximation of the dipole field by

\begin{equation}
\vartheta_c \simeq \left({a\omega \over c}\right)^{1/2},
\label{thetc}
\end{equation}

where $\omega $ is the angular velocity of  a rotating pulsar and $c$
is the speed of light.

The polar zone, which does not corotate with the pulsar,
is active since it is subject to an electromotive force (emf).
Similarly as for an unipolar inductor the value of the emf or
the difference of an electrostatic
potential, $\varphi$, between the pole and a point at the latitude
$\vartheta_c$ can be found as follows:
\begin{equation}
\Delta \varphi = E\ell \simeq {a^3 B_0 \omega ^2 \over c^2},
\end{equation}
where the electric field relates to the magnetic field
at the surface of the star, $B_0$, through
\begin{equation}
E \simeq {\ell \omega \over c}B_0
\end{equation}
and the characteristic length is
\begin{equation}
\ell = \vartheta_c a .
\end{equation}

Goldreich and Julian (1969) assumed that the pulsar is isolated and
showed that if the electric, {\bf E}, and  magnetic, {\bf B}, fields
fulfil the condition

\begin{equation}
{\bf E \cdot B}=0
\end{equation}
at least approximately, a net charge density, to be
expected, is given by

\begin{equation}
|\rho_c| \simeq { \omega B_0 \over 2\pi c} .
\end{equation}
Assuming that the currents are due to a completly charge separated
plasma, the current density is approximately given by

\begin{equation}
J=|\rho_c| c .
\label{curc}
\end{equation}
One can than evaluate the total current

\begin{equation}
I \approx \pi\ell ^2 J \simeq {B_0 a^3 \omega ^2 \over 2 c}
\label{curt}
\end{equation}
and the energy loss rate, which reads

\begin{equation}
\left({dW \over dt}\right)^P =I \Delta \varphi \simeq
{\omega^4 a^6 B_0^2 \over 2 c^3} .
\label{power}
\end{equation}
This is the formula proposed by Goldreich and Julian (1969), which
substantially coincides with the radiating dipole formula of
Deutsch (1955).


A rather controversial and specific assumption of the Goldreich
and Julian paper is that of complete charge separation. However
it is noticeable that also in models where the currents are almost
neutral, eq. (7) appears to be valid. In particular
Beskin et al. (1993) has shown that if the plasma pressure
is insignificant and the density of positive or negative charges, $|\rho|$,
satisfies the following condition
\begin{equation}
|\rho| \gg |\rho_c|,      \label{rho}
\end{equation}

the Grad-Shafranov
equation combined with the boundary conditions for the active region can
be explicitly solved. The important conclusion is that eqs  (7)-(9) are
recovered. In any case we assume in the following that (7) holds
as order of magnitude.

\section{ Accretion onto a magnetized neutron star}

Consider a non-rotating neutron star endowed of a magnetic field
$B_0$ accreting a homogeneous medium at a rate $\dot M$. The
associated luminosity is

\begin{equation}
\left({dW \over dt}\right)^A = {GM\dot M \over a} .
\end{equation}
The simplest picture asserts that the magnetic field can be
neglected at radii larger than the Alfv\`en radius

\begin{equation}
r_A =\left({2 \mu^2 \over \dot M \sqrt{2GM}}\right)^{2/7},
\end{equation}
where $\mu ={1 \over 2} a^3 B_0$.
This is the distance where the magnetic energy density
approximately equals the energy density of the infalling medium.
The critical magnetic surface of equatorial radius, $r_A$,
intersects  the neutron star surface at a colatitude

\begin{equation}
\vartheta _A = \left({a \over r_A}\right)^{1/2}.
\end{equation}

Supposedly matter  moves along the critical magnetic surface
filling completely the magnetic funnels
and hits the neutron star at two polar caps  which have a
radial extension

\begin{equation}
r=a\vartheta _A ={ a ^{3/2} \over r_A ^{1/2}}.
\end{equation}
If centrifugal force at the Alfv\`en radius overcomes gravitation,
accretion may be inhibited (propeller phase). However we  consider
here only the situation where accretion does occur, since we are
interested in the short circuiting of the polar cap of the neutron star.

\section{ Electrodynamics of a rotating magnetized accretor}

Suppose that the neutron star is rotating with the spin axis
parallel to the magnetic axis, and that it is accreting at a rate
$\dot M$. Further suppose that

\begin{equation}
\vartheta _A > \vartheta _c.
\end{equation}

The region enveloped by the magnetic surface of colatitude
$\vartheta_A $  will be
electrodynamically inactive. In fact it corotates with the neutron
star and no emf is present. Moreover at least in the simplest
picture, the accreting plasma cannot penetrate this region. The regions
of electrodynamical activity will be those corresponding to the
accretion funnels within colatitude
$\vartheta_A$. The accreting material does not corotate with
the star
and its density dominates that of the plasma produced by the
neutron star rotation. Therefore the case is closer to that treated
by the Beskin, Gurevich and Istomin (1993) rather
than to the original Goldreich and Julian one. The main difference
is that the boundary of the active region is now  $\vartheta_A$
and not $\vartheta_c$. It follows that the relevant equation
for the electrodynamics will be again (\ref{curc}), since it is
unaffected  by
the boundary condition for the inactive
region. The equations for the emf, current and power are recovered
substituting the characteristic length, $ \ell$,
by $r$ from  equation (14),
and therefore one has

\begin{equation}
\Delta^A\varphi \simeq {\omega B_0 \over c }r^2,
\end{equation}

\begin{equation}
I^A \simeq {\omega B_0 r^2\over 2},
\end{equation}
and

\begin{equation}
\left({dW \over dt}\right)^{AP}=I^A \Delta ^A \varphi \simeq
{(\omega B_0)^2r^4 \over 2 c}=
{2 \omega \mu^2 \over r_A^2 r_L},
\end{equation}
where

\begin{equation}
r_L={c \over \omega}.
\end{equation}
The corresponding torque  is given by

\begin{equation}
K^{AP} = {2\mu^2 \over r_A^2 r_L}  \label{kap}
\end{equation}

Eqs (18) and (20) are in our opinion reasonable expressions for
the energy and angular momentum loss associated to currents reaching the
neutron star.  Obviously they are not the result of a self consistent
picture. The actual current structure remains an exceedingly complicated
and still unsolved problem.

\section{Some astrophysical consequences}

In order to explore the astrophysical consequences of our scenario
of the electrodynamics of accreting magnetized rotators we
examine two cases which may be representative for X-ray binaries.
In both the mass of a neutron star is taken equal to 1 $M_{\odot}$
and its radius to $10^6$ cm.
First we examine the case of Her
X-1, a prototype of X-ray pulsators. We take
$ B_0 \simeq 10^{12}$ G, $P=1.2$ s, and $\dot M=10^{17}$ g/s.
{}From the previous formulae one has

\begin{equation}
\left({dW \over dt}\right)^A \sim 10^{37} \rm erg/s
\end{equation}

\begin{equation}
\left({dW \over dt}\right)^{AP} \sim  10^{34} \rm erg/s
\end{equation}

\begin{equation}
K^{AP} \sim  10^{33} \rm erg
\end{equation}
The second case is that of luminous low mass X-ray binaries. Here
we take $\dot M =10^{18}$ g/s.
The values of the magnetic field and period are  unknown.
We will consider $B_0\simeq 10^9$ G and a period of 5 milliseconds,
and scale accordingly. The result is

\begin{equation}
\left({dW \over dt}\right)^A  \sim 10^{38} \rm erg/s
\end{equation}

\begin{equation}
\left({dW \over dt}\right)^{AP} \sim
 10^{36} B_9^{6/7}P_5^{-2} \rm erg/s
\end{equation}

\begin{equation}
K^{AP} \sim 10^{33} \rm erg
\end{equation}

It is apparent that in both cases the energy released by the
passage of  currents is a small fraction ($10^{-2}-10^{-3}$) of the
accretion luminosity. It should be pointed out here that
the interaction between
the magnetosphere and infalling material,  which is not
included here, may affect this results.
Our picture is obviously inadequate for
spectral calculations. However one
could argue that the current structure may build up a non thermal
contribution. In particular it is tempting to apply this scheme
to a system like Cyg X-3, where rotation of a rapid pulsar may
coexist with a large  accretion rate. The resulting model would
then have close relation to the flywheel picture of the system
(Treves and Bocci, 1987, Mineshinge, Fabian, Rees, 1991).

The other aspect where this picture may be relevant is that related
on the torque acting on the neutron star (\ref{kap}). The torque
is always a braking one and therefore tends to spin down the
pulsar. Spin up and spin down on accreting neutron stars have
been the focus of a diffuse interest, because they are directly related
to the observed period derivative of X-ray pulsators.
The theory is somewhat intricated and we refer for a recent presentation
to Lipunov (1992). The spin up is supposedly dominated by the
angular momentum transferred from the accretion disk at the last
Keplerian orbit, and depends on the accretion rate. Here we focus
on the spin down. In the usual picture the spin down torque is
calculated through the emergence of a toroidal component of the
magnetic field in the accretion disk (e.g. Lynden-Bell and Pringle 1974).
Alternatively one can consider viscous friction between the plasma
and the magnetosphere (e.g. Davies and Pringle 1981). Both approaches
yield the formula (Lipunov 1992)


\begin{equation}
K^L={\mu^2 \over r_c^3}  \label{kl}
\end{equation}
where $r_c$ is the corotation radius

\begin{equation}
r_c=\left({GM \over \omega ^2}\right)^{1/3}
\end{equation}
Equation (\ref{kl}) is formally similar to our equation (\ref{kap})
where $r_A^2r_c$  plays the role of $r_c^3$.
Applying this expression to the two examined cases, we obtain for
Her X-1

\begin{equation}
K^L \sim  10^{34} \rm erg
\end{equation}
and for the second representative case

\begin{equation}
K^L \sim  10^{33} \rm erg
\end{equation}

In the first case the torque considered by us is one order of
magnitude lower than that given by (\ref{kl}). In the second case
the torques  are comparable. Since

\begin{equation}
{K^{AP} \over K_L} = {r_c^3 \over r_A^2 r_L} \sim {M^{9/7}
\dot M^{4/7} \over \omega \mu^{8/7}}
\end{equation}
it is clear that if the magnetic field of a considered LMXRB
were smaller than $10^9$ G the  torque proposed here might even
exceed the torque
described by (27).
This indicates that
the braking corresponding to
equation
(\ref{kap}) should not
be neglected.

{\bf Acknowledgements}

We are grateful to M. Camenzind and V. Lipunov for fruitful discussion

{\bf References}

Beskin, V. S., Gurevich, A. V., and Istomin, Ya., N., 1993, {\it Physics
of the Pulsar Magnetosphere}, Cambridge University Press 1993

Davies, R. E., and Pringle, J. E., 1981, {\it M.N.R.A.S.} {\bf 196}, 209

Deutsch, A. J., 1955, {\it Annales d'Astrophysique} {\bf 18}, 1

Goldreich, P., and Julian, W. H., 1969, {\it Ap. J} {\bf 157}, 869

Lipunov, V. M., 1992, {\it  Astrophysics of Neutron Stars},
Springer-Verlag Berlin Heidelberg 1992

Lynden-Bell D., and Pringle, J. E., 1981, {\it M.N.R.A.S} {\bf 168}, 603

Mineshige, S, Rees, M. J., and Fabian, A. C., 1991, {\it M.N.R.A.S.}
{\bf 251}, 555

Treves, A., Bocci F., 1987, {\it M.N.R.A.S}, {\bf 225}, 39P

\end{document}